\documentclass[journal]{IEEEtran}

\ifCLASSINFOpdf
\else
   \usepackage[dvips]{graphicx}
\fi
\usepackage{url}

\usepackage{graphicx}
\usepackage{amsmath}
\usepackage{balance}
\usepackage{algorithm}
\usepackage{algpseudocode}

\begin{document}

\title{SSCFormer: Push the Limit of Chunk-wise Conformer for Streaming ASR Using Sequentially Sampled Chunks and Chunked Causal Convolution}

\author{Fangyuan Wang, \IEEEmembership{Member, IEEE}, Bo Xu, Bo Xu, \IEEEmembership{Member, IEEE}
\thanks{This paragraph of the first footnote will contain the date on which you submitted your paper for review. For example, ``This work was supported in part by the U.S. Department of Commerce under Grant BS123456.'' }
\thanks{The next few paragraphs should contain the authors' current affiliations, including current address and e-mail. }
\thanks{S. B. Author, Jr., was with Rice University, Houston, TX 77005 USA. He is now with the Department of Physics, Colorado State University, Fort Collins, CO 80523 USA (e-mail: author@lamar.colostate.edu).}}

\markboth{Journal of \LaTeX\ Class Files, Vol. 14, No. 8, August 2015}
{Shell \MakeLowercase{\textit{et al.}}: Bare Demo of IEEEtran.cls for IEEE Journals}
\maketitle

\begin{abstract}
Currently, the chunk-wise schemes are often used to make Automatic Speech Recognition (ASR) models to support streaming deployment. However, existing approaches are unable to capture the global context, lack support for parallel training, or exhibit quadratic complexity for the computation of  multi-head self-attention (MHSA).
On the other side, the causal convolution, no future context used, has become the
\emph{de facto} module in streaming Conformer. 
In this paper, we propose SSCFormer to push the limit of chunk-wise Conformer for streaming ASR using the following two techniques: 1) A novel cross-chunks context generation method, named Sequential Sampling Chunk (SSC) scheme, to re-partition chunks from regular partitioned chunks to facilitate efficient long-term contextual interaction within local chunks. 2)The Chunked Causal Convolution (C2Conv) is designed to concurrently capture the left context and chunk-wise future context. Evaluations on AISHELL-1 show that an End-to-End (E2E) CER 5.33\% can achieve, which even outperforms a strong time-restricted baseline U2. 
Moreover, the chunk-wise MHSA computation in our model enables it to train with a large batch size and perform inference with linear complexity
\footnote{We will release the code in the future.}.
\end{abstract}

\begin{IEEEkeywords}
conformer, streaming ASR, sequentially sampled chunks, chunked causal convolution, linear complexity 
\end{IEEEkeywords}

\IEEEpeerreviewmaketitle

\section{Introduction}
\IEEEPARstart{I}{n} recent times, Conformer models~\cite{2020_conformer,Zhang2020UnifiedSA} that combine self-attention~\cite{NIPS2017_transformer} with convolution, have exhibited excellent performance and outperformed Connectionist Temporal Classification (CTC) ~\cite{li2018advancing,Graves2006ConnectionistTC}, Recurrent Neural Network Transducer (RNN-T) \cite{Battenberg2017ExploringNT}, and Transformer~\cite{Battenberg2017ExploringNT} models in numerous Automatic Speech Recognition (ASR) tasks~\cite{Li2020OnTC,Guo2021RecentDO}.

Inspired by the remarkable performance of Conformer models, researchers aim to modify them for streaming ASR. The primary focus is on developing a Conformer encoder for online ASR that integrates streaming self-attention and convolution. Two primary approaches for streaming self-attention are chunk-wise self-attention and causal self-attention, as described in~\cite{An2022CUSIDECS} and~\cite{Moritz2021DualCS}. Generally, chunk-wise encoders exhibit better accuracy than causal encoders since they utilize future features, which the latter do not~\cite{An2022CUSIDECS}.

\begin{figure}[htb]
\begin{minipage}[b]{1.0\linewidth}
  \centering
  \centerline{\includegraphics[width=8.5cm]{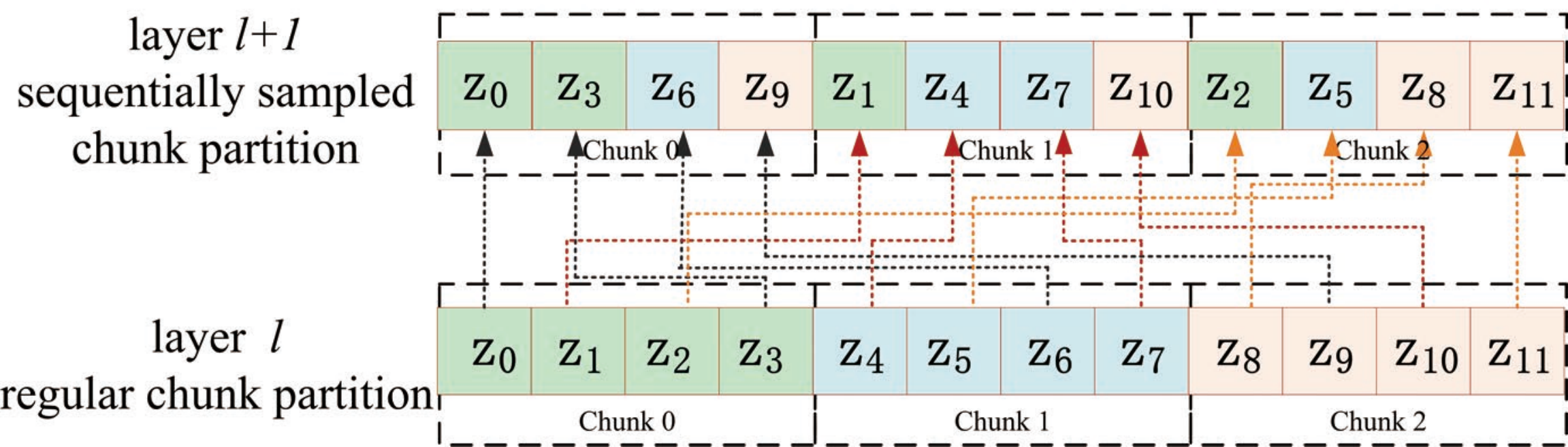}}
%  \vspace{2.0cm}
\end{minipage}
\caption{
An illustrative example of sequential sampling chunk partition scheme.
In layer \emph{l}, MHSA is computed in regular chunks. However, in the next layer (\emph{l}+1), MHSA is computed in sequentially sampled chunks that enable cross-chunk interactions for tokens. $z_i$, $i \in \{0,1,..,11\}$, denotes a speech token. }
\label{fig:ssc-chunk}
\end{figure}

The regular chunk-wise methods, as described in~\cite{Tian2019SynchronousTF,Wang2020ReducingTL,Chen2021DevelopingRS,Li2021HeadSynchronousDF}, compute MHSA locally in chunks, as illustrated in layer $l$ of Fig.~\ref{fig:ssc-chunk}. Despite their linear complexity, these methods suffer substantial performance degradation due to their inability to model cross-chunk context.
Typically, the time-restricted methods ~\cite{Zhang2020UnifiedSA,An2022CUSIDECS,Moritz2020StreamingAS,Yu2020UniversalAU,Wu2021U2UT,Wang2021WNARSWB,Weninger2022ConformerWD} can significantly alleviate the degradation by using the concatenation of current and previous chunks to perform MHSA computation. However, these models have quadratic complexity due to global MHSA, which may lead to unacceptable latency for long inputs.
Some researchers have attempted to enhance modeling capabilities while maintaining linear complexity. For instance, the memory-based chunk-wise methods~\cite{Wu2020StreamingTA,Zhang2020StreamingCM,Inaguma2020EnhancingMM,Shi2021EmformerEM} utilize the memory mechanism to incorporate historical context, along with local chunk-wise MHSA. This approach can improve modeling capabilities, but it comes at the expense of violating the parallel training property.
A recent work~\cite{Wang2022ShiftedCE} presented a streaming Conformer encoder that significantly enhances accuracy by enabling cross-chunk interactions via a shifted chunk scheme, outperforming regular chunk-wise methods.
Nonetheless, the half-shifted chunks used in this approach are inefficient in capturing long-term context since they only allow MHSA computation across adjacent chunks.
Besides, most streaming Conformer models~\cite{Zhang2020UnifiedSA,An2022CUSIDECS,Wu2021U2UT,Weninger2022ConformerWD,Wang2022ShiftedCE} rely on causal convolution to suit the streaming deployment and can only utilize the left context.

This paper introduces a novel streaming ASR model called SSCFormer, which stands for \textbf{S}equentially \textbf{S}ampled \textbf{C}hunk Con\textbf{former}. 
Its core contributions include:

\begin{itemize}
\item[1)] 
Using Sequentially Sampled Chunks to enable chunk-wise MHSA to effectively capture long-term contexts within linear complexity, see layer $l+1$ in Fig.~\ref{fig:ssc-chunk}; 

\item[2)]
Incorporating a Chunked Causal Convolution to capture history and chunk-wise future features, see Fig.~\ref{fig:c2conv}. 
\end{itemize}

Experiments on AISHELL-1 show that it can achieve CER of 5.33\%, surpassing the U2\cite{Zhang2020UnifiedSA} with quadratic complexity.

\section{SSCFormer}

\subsection{Overall architecture}
\label{ssec:CC}
Our SSCFormer model follows the same framework as U2~\cite{Zhang2020UnifiedSA}, which includes a Conformer encoder and a CTC/Attention hybrid decoder. We utilize the same decoder and update the encoder using our proposed techniques. The updated encoder incorporates SpecAug~\cite{Park2019SpecAugmentAS}, SpecSub~\cite{Wu2021U2UT}, and other frontend layers, followed by consecutive Chunk-C2Conv Conformer blocks and SSC-C2Conv Conformer blocks, as illustrated in Fig.~\ref{fig:overal_arch}(a), to replace the vanilla Conformer blocks.

\subsection{Chunk-C2Conv/SSC-C2Conv Conformer block}
\label{ssec:ssc-MHSA}
\begin{figure}[htb]
\begin{minipage}[b]{1.0\linewidth}
  \centering
  \centerline{\includegraphics[width=5.2cm]{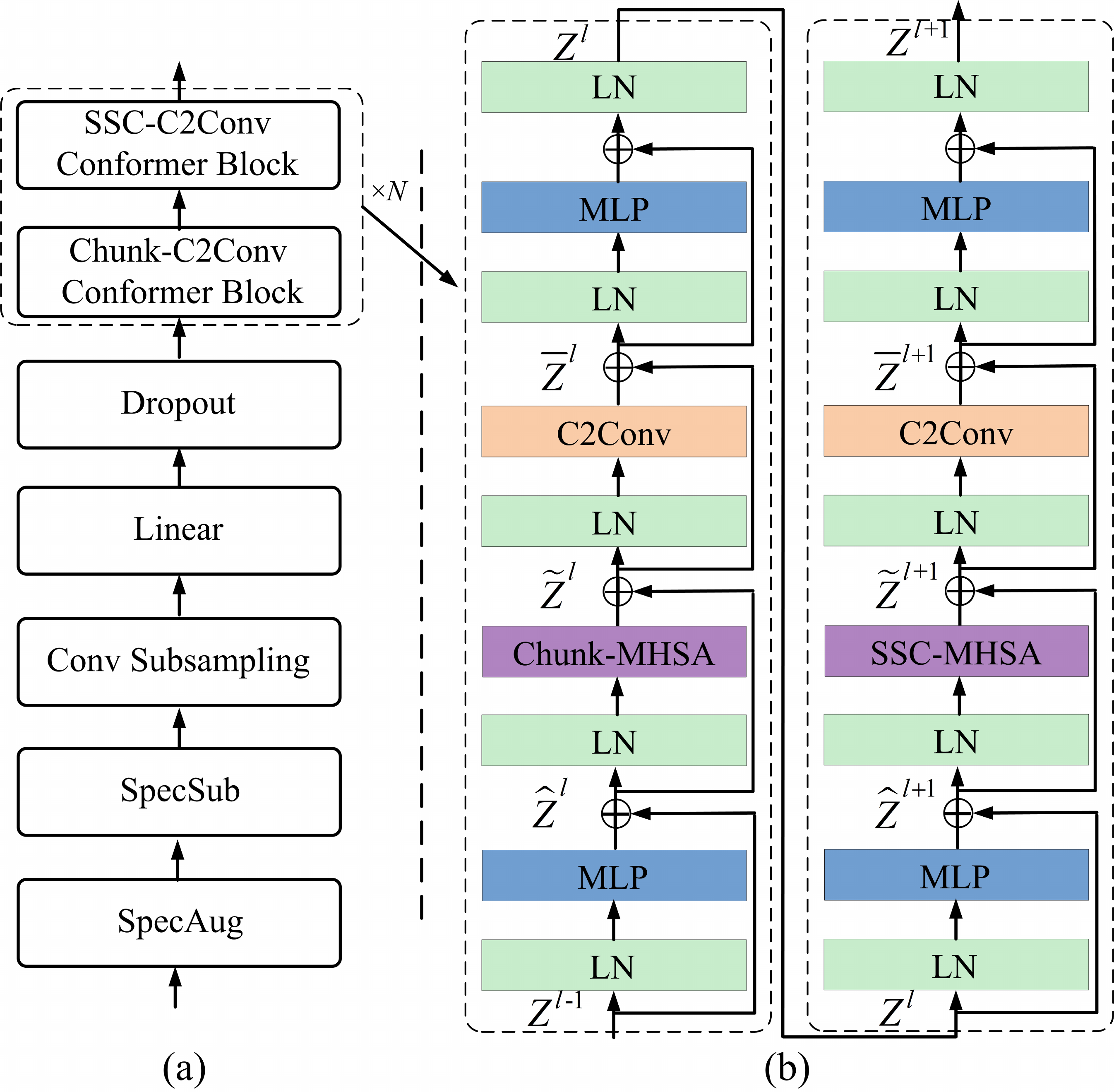}}
  %\vspace{0.1cm}
\end{minipage}
\caption{(a) The overview of the SSCFormer encoder; (b) two successive Chunk-C2Conv Conformer and SSC-C2Conv Conformer blocks (\emph{N}=6, notations see in Section II.B).
}
\label{fig:overal_arch}
\end{figure}

To construct the Chunk/SSC-C2Conv Conformer block, we replace the MHSA and the causal convolution with Chunk/SSC-MHSA (as described in Section II.C and II.D) and C2Conv (as explained in Section II.E), respectively, as shown in Fig.~\ref{fig:overal_arch}(b). This block comprises two Multi-Layer Perceptron (MLP) surrounding Chunk/SSC-MHSA and C2Conv. Prior to each MHSA, MLP, and C2Conv, it applies LayerNorm (LN) and and incorporates a residual connection after each module.
The computation of successive blocks is as follows:

\begin{equation}
	\begin{aligned}	 
    \hat{Z}^{l} &= 0.5\times{MLP(LN(Z^{l-1}))}+Z^{l-1}\text{,}\\
    \tilde{Z}^{l}&=Chunk\text{-}{MHSA}(LN(\hat{Z}^{l}))+\hat{Z}^{l}\text{,}\\
    \overline{Z}^{l} &=C2Conv(LN(\tilde{Z}^{l}))+\tilde{Z}^{l}\text{,}\\
    Z^{l} &=LN(0.5\times{MLP(LN(\overline{Z}^{l}))}+\overline{Z}^{l}),\\
    \hat{Z}^{l+1} &= 0.5\times{MLP(LN(Z^{l}))}+Z^{l}\text{,}\\
    \tilde{Z}^{l+1}&=SSC\text{-}{MHSA}(LN(\hat{Z}^{l+1}))+\hat{Z}^{l+1}\text{,}\\
    \overline{Z}^{l+1} &=C2Conv(LN(\tilde{Z}^{l+1}))+\tilde{Z}^{l+1}\text{,}\\
     Z^{l+1} &=LN(0.5\times{MLP(LN(\overline{Z}^{l+1}))}+\overline{Z}^{l+1})\\
    \end{aligned}\label{eq1}
\end{equation}
where $\hat{Z}^{l}$, $\tilde{Z}^{l}$, $\overline{Z}^{l}$, and $Z^{l}$ are the outputs for the first MLP, Chunk/SSC-MHSA, C2Conv, and the second MLP of layer \emph{l};
C2Conv is the convolution module we used; Chunk/SSC-MHSA refers to chunk-wise MHSA, using regular and sequential sampling chunk partitioning contexts, respectively.

\subsection{Regular partitioned chunks based MHSA}
The calculation of Chunk-MHSA is within regular partitioned chunks to capture chunk-wise contexts, e.g. \emph{layer l} in Fig.~\ref{fig:ssc-chunk}, where the input sequence is evenly divided into non-overlapped chunks.
Typically, this regular chunk-wise MHSA does not use any long-term contextual information.

\subsection{Sequentially sampled chunks based MHSA}
\textbf{Sequential sampling chunk partition.}
To improve the modeling capacity of standard chunk-wise MHSA, we introduce a partitioning approach that enables cross-chunk interactions by sequentially sampling tokens from regular chunks and consolidating them into new ones.
Fig.~\ref{fig:ssc-chunk} gives an example. We sample from the beginning in layer \emph{l} with interval 3 (the number of chunks), resulting in the first chunk [$z_0$, $z_3$, $z_6$, $z_{9}$] in layer \emph{l+1}.
Then we move forward in layer \emph{l} and repeat the above process
until all tokens have been relocated.
The MHSA,  computed within the resulting new chunks, can capture the long-range contextual features efficiently.

\begin{figure*}[!t]
\begin{minipage}[b]{1.0\linewidth}
  \centering
  \centerline{\includegraphics[width=6.4in]{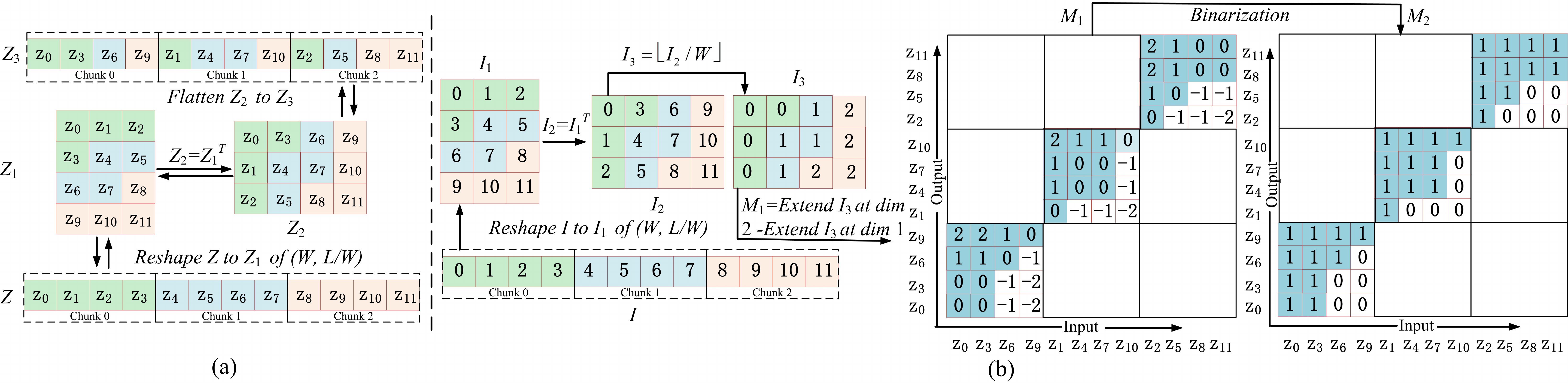}}
%  \vspace{2.0cm}
\end{minipage}
\caption{Illustration of (a) the efficient batch generation of sequentially sampled chunks and (b) the dynamic generation of attention mask for sequentially sampled chunks. In this example, the sequence length and chunk size are denoted by \emph{L} and \emph{W}, and are set to 12 and 4, respectively. }
\label{fig:ssc-chunk-para}
\end{figure*}

\textbf{Efficient batch computation for sequential sampling.}
We need to pad audios in a batch as in \cite{Wang2022ShiftedCE} and convert the sequential sampling operations to matrix computations to support parallel processing, see the steps in Fig.~\ref{fig:ssc-chunk-para}(a).

\begin{figure}[htb]
\begin{minipage}[b]{1.0\linewidth}
  \centering
  \centerline{\includegraphics[width=8.5cm]{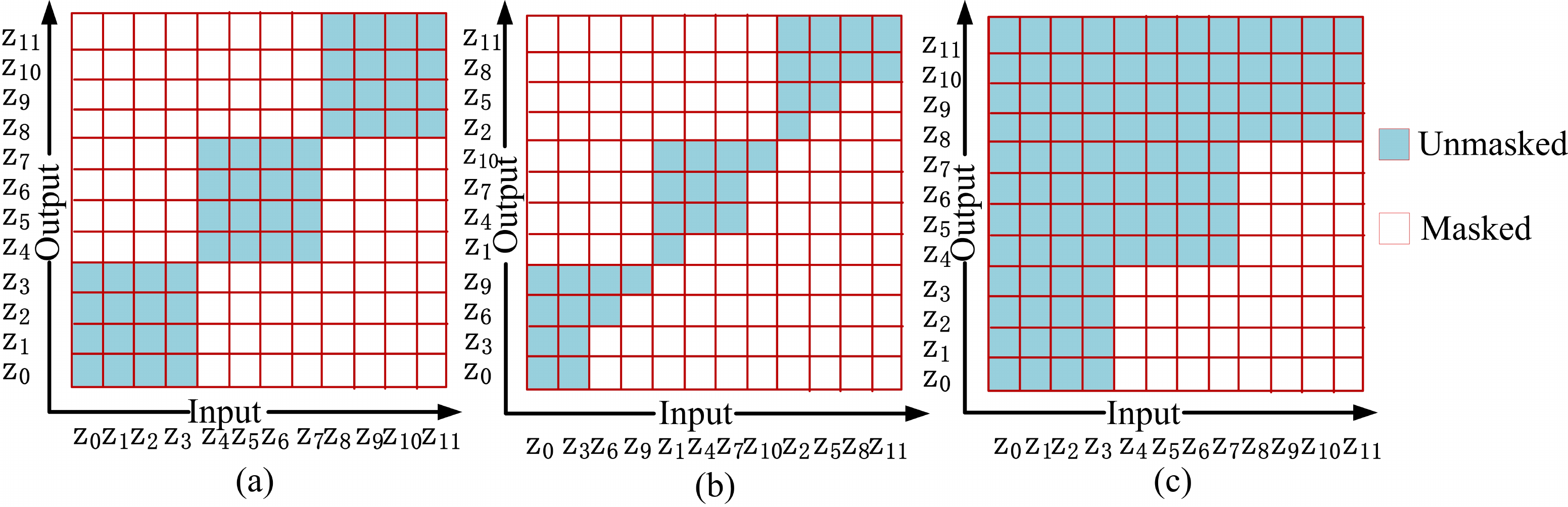}}
%  \vspace{2.0cm}
\end{minipage}
\caption{Attention masks of 3 chunks with chunk size 4 for (a) regular Chunk-MHSA, (b) SSC-MHSA, and (c) time-restricted MHSA. }
\label{fig:ssc_chunk_mask}
\end{figure}

\textbf{Sequentially sampled chunks attention mask.}
The Chunk/SSC-MHSA needs attention mask to avoid interactions with forbidden future tokens under streaming configuration.
The distinctiveness of the SSC-MHSA attention mask is that they are generated dynamically, 
Fig.~\ref{fig:ssc-chunk-para}(b) gives an illustration of its generation process.
The attention masks for Chunk-MHSA, SSC-MHSA, and time-restricted MHSA are comparatively illustrated in Fig.~\ref{fig:ssc_chunk_mask}(a,b,c), respectively.

\textbf{Complexities of MHSA operations\footnote{
We omit the softmax when determining complexity. \emph{Q}, \emph{K}, and \emph{V} represent the matrices of query, key, and value in Transformer, respectively.
}.}
Suppose \emph{L} is the input length and \emph{C} is the feature dimension.
The complexity of global MHSA = $\Omega(MHSA)$ = $3L{C^2}$ (linear transformations of \emph{Q}, \emph{K}, and \emph{V}) 
+ $L^2{C}$ (dot product of $QK^T$) 
+ $L^2{C}$ (the transformation of \emph{V} multiples attention scores) 
+ $L{C^2}$ (the final linear transformation) 
= $4L{C^2}+2L^2{C}$.
The MHSA calculations are performed locally. If \emph{W} denotes the chunk size, the complexity of chunk-wise MHSA can be expressed as $\Omega(Chunk/SSC\text{-}MHSA) = L/W\left(4W{C^2}+2W^2{C}\right) = 4L{C^2}+2W{L}{C}$, which is linear when \emph{W} is constant.
In contrast, the time-restricted methods perform MHSA over all unmasked chunks, as shown in Fig.~\ref{fig:ssc_chunk_mask}(c), resulting in an average complexity that is half that of global MHSA.

\subsection{Chunked causal convolution}
\label{ssec:CC}
\begin{figure}[htb]
\begin{minipage}[b]{1.0\linewidth}
  \centering
  \centerline{\includegraphics[width=8.0cm]{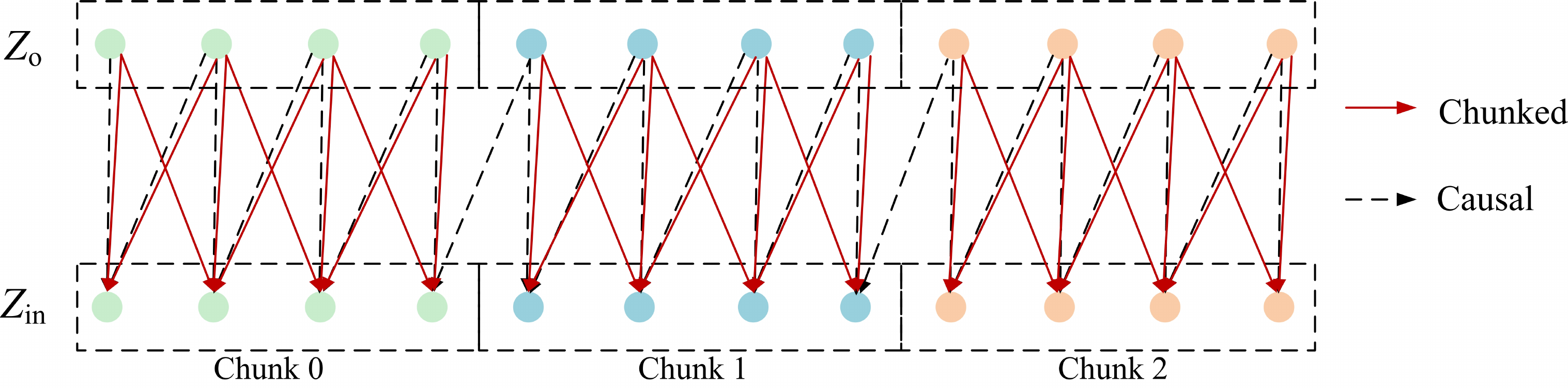}}
%  \vspace{2.0cm}
\end{minipage}
\caption{Illustration of the computation process of Chunked Causal Convolution, where the kernel size of masked convolution is 3 and the chunk size is 4. The causal convolution can cross chunks, while the chunked convolution can attend to chunk-wise future tokens.}
\label{fig:c2conv}
\end{figure}

In Figure~\ref{fig:c2conv}, we present the Chunked Causal Convolution, which combines the cross-chunk left context captured by the causal convolution and the chunk-wise future feature captured by the chunked convolution in the convolution layer.
The Chunked Causal Convolution utilizes a parameter-sharing module that is implemented through a masked convolution. In the causal convolution, future tokens in the kernel are masked; however, in the chunked convolution, they are not masked. The chunked convolution and causal convolution are computed separately, and their outputs are combined as follows:
$Z_o = \lambda \times Chunked\text{-}Conv(Z_{in}) + (1-\lambda) \times Causal\text{-}Conv(Z_{in})$,
where $\lambda$ is a hyperparameter, and $Z_{in}$ and $Z_o$ represent the input and output vectors, respectively. As the MHSA layers use the same chunk size, this approach does not introduce any extra latency.

\subsection{Streaming inference}
In the SSCFormer encoder, the inputs are consumed chunk by chunk. 
In the case of the first chunk \{$z_0$, $z_1$, $z_2$, $z_3$\}, there is no need to perform sequential sampling when computing the attention, and the SSC-MHSA layer behaves like the Chunk-MHSA layer without introducing any additional delay.
For following chunks, the encoder caches the previous chunks and concatenates them with the current chunk before processing. 
This approach enables the model to interact with tokens from historical chunks while preserving a chunk-wise processing strategy.
For instance, for a given sampled chunk \{$z_0$, $z_3$, $z_6$, $z_9$\}, as illustrated in Fig.~\ref{fig:ssc_chunk_mask}(b), $z_0$ and $z_3$ can attend to $z_0$ and $z_3$ because $z_6$ and $z_9$ have been masked out, $z_6$ can attend to $z_0$, $z_3$, and $z_6$, while $z_9$ can attend to all the tokens within the sampled chunk.
Upon receiving the encoder output, the CTC decoder generates the initial hypotheses without delay. Once an utterance is completed, the Attention decoder is employed to reevaluate the output of the CTC decoder and generate more precise results at the utterance level.
\section{Experiments}
\label{sec:experiments}

\subsection{Data}
We perform evaluations on the AISHELL-1 dataset~\cite{Bu2017AISHELL1AO}, which comprises a training set of 150 hours, a dev set of 10 hours, and a test set of 5 hours, totaling 7176 utterances.

\subsection{Experimental Setup}
All models are implemented using the WeNet 2.0 toolkit~\cite{zhang2022wenet} and evaluated on two RTX 3090 GPUs (24G).
We largely adopted the hyperparameter settings suggested by WeNet, such as using FBANK features 80-dimensional computed on-the-fly by torch audio with a 25-ms window and a 10-ms shift.
Here are the distinctive hyperparameters that we used: 1) The kernel size of the C2Conv layer is set to 15, with 7 right tokens masked for the causal convolution and none for the chunked convolution. 2) The encoder comprises 6 interleaved Chunk-C2Conv Conformer blocks and 6 SSC-C2Conv Conformer blocks. 3) The final model is obtained by taking the average of the top 30 checkpoints with the best validation loss. 4) To utilize GPU memory effectively and speed up training, we set the batch size to 48 for chunk-wise models and 36 for time-restricted models.

\subsection{Ablation studies}
Firstly, we evaluate the effectiveness of our proposed sequentially sampled chunks and chunked causal convolution over various chunk sizes.
Table~\ref{tab:chunk_size} lists the results. 
Comparing C.1 with B and A, we observe that models using our proposed sequential sampling chunk partition scheme can consistently reduce the CERs across various chunk sizes. This suggests that our SSC scheme can effectively capture cross-chunk context more effectively than regular and shifted chunk schemes.
Additionally, comparing C.2 with C.1 confirms that integrating the chunk-wise right context and the causal left context is beneficial in further improving the overall accuracy.
In order to achieve a balance between accuracy and latency, we set the chunk size to 16 in our subsequent experiments.
Our second investigation is about the chunked convolution weight $\lambda$, see Table~\ref{tab:lambda}.
We observe that the SSCFormer with the integration of causal convolution and chunked convolution can outperform that only used them individually. The best CER of 5.33 was achieved when setting $\lambda$ to 0.7, including the results in Table~\ref{tab:chunk_size}.

\begin{figure}[htb]
\begin{minipage}[b]{1.0\linewidth}
  \centering
  \centerline{\includegraphics[width=5.0cm]{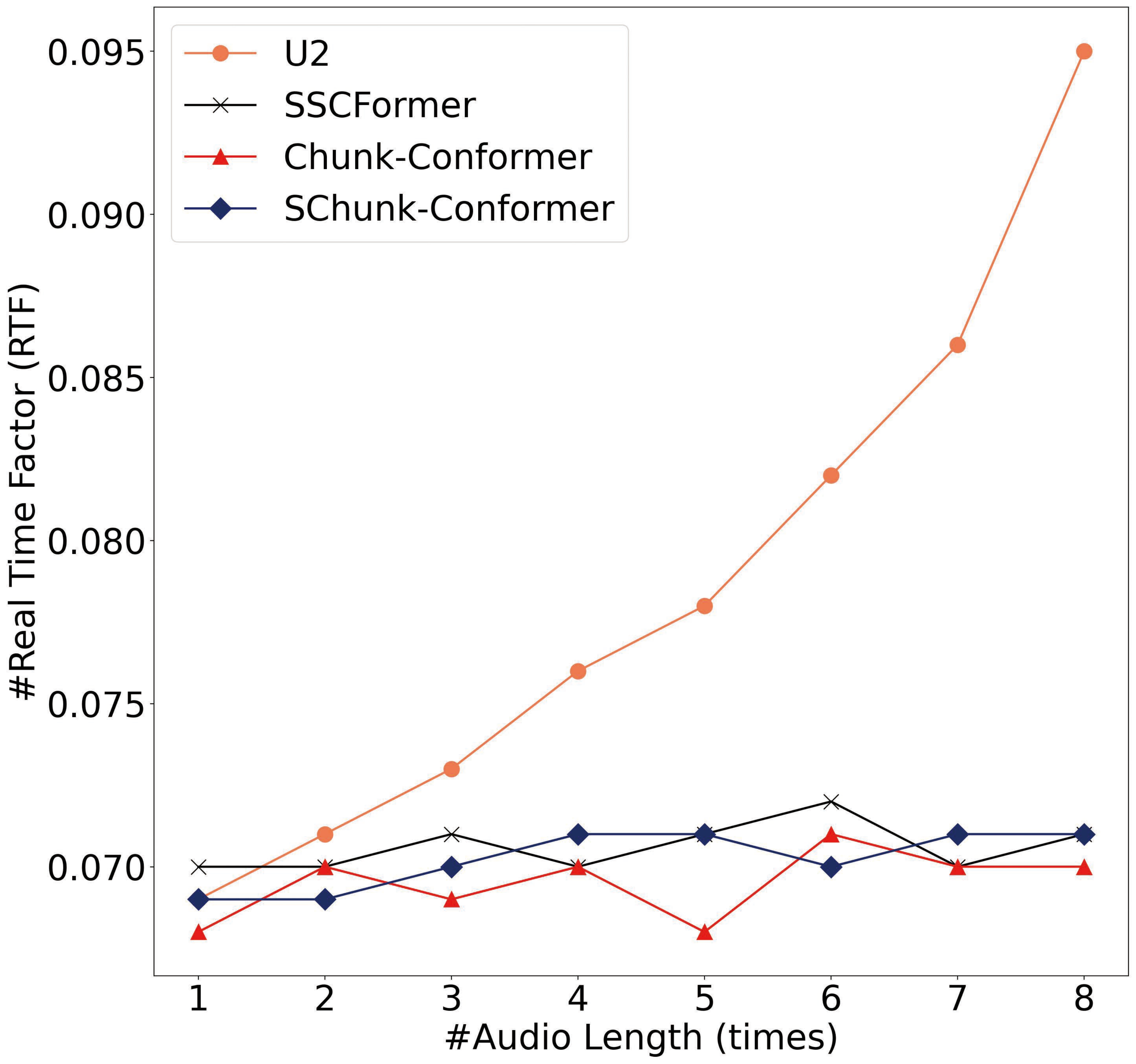}}
%  \vspace{2.0cm}
\end{minipage}
\caption{The comparison of RTFs for SSCFormer with other models.
To simulate different audio lengths, we concatenate each audio with itself several times in the AISHELL-1 test set. All evaluations are performed on a single CPU (Intel(R) Xeon(R) Silver 4210R CPU @ 2.40GHz) using a single thread.
}
\label{fig:time_cost}
\end{figure}

\begin{table}[t]
\centering
\setlength{\tabcolsep}{0.4mm}
\caption{ Ablation study of the SSCFormer architecture (CER\%)}
\label{tab:chunk_size}
\begin{tabular}{|c|c|c|c|c|c|c|c|c|}
\hline
  &\# Chunk Size & Partition & Conv. &4 &8 &10 &16 &20\\ 
\hline
A & Chunk-Conformer (repro) &Regular & Causal &$6.55$ &$6.33$ &$6.30$ &$6.09$ &$6.01$\\
B & SChunk-Conformer~\cite{Wang2022ShiftedCE} &Shifted &Causal &$6.74$ &$6.21$ &$6.09$ &$5.77$ &$5.72$\\
\hline
C.1 &\textbf{SSCFormer (w/o C2Conv)} &SSC & Causal &\textbf{6.45} &\textbf{6.08} &\textbf{5.97} &\textbf{5.58} &\textbf{5.48}\\
C.2 &\textbf{SSCFormer (w/ C2Conv)} &SSC & C2Conv &\textbf{6.41} &\textbf{5.91} &\textbf{5.76} &\textbf{5.33} &\textbf{5.26}\\
\hline
\end{tabular}
\end{table}

\begin{table}[t]
\centering
\setlength{\tabcolsep}{1.0mm}
\caption{ Effect of chunked convolution weight $\lambda$}
\label{tab:lambda}
\begin{tabular}{|c|c|c|c|c|c|c|c|c|c|c|c|}
\hline
  $\lambda$ &\ 0 & 0.1 & 0.2 &0.3 &0.4 &0.5 &0.6 &0.7 & 0.8 & 0.9, & 1.0\\ 
\hline
CER(\%) &5.58 &5.54 & 5.51 &5.43 &5.39 &5.42 &5.37 &\textbf{5.33} &5.35 &5.44 & 6.19\\
\hline
\end{tabular}
\end{table}

\subsection{Comparisons with other models}

\begin{table}[t]
\centering
\setlength{\tabcolsep}{1.8mm}
\caption{ Training efficiency comparisons. 
The maximum batch size is denoted as "max. batch (\#)", while "trn. time (h)" refers to the training time for models trained over 180 epochs.}
\label{tab:baseline}
\begin{tabular}{|c|c|c|c|}
\hline
Model &Type &Max. &Trn. \\
Architecture & &Batch (\#)  &Time (h)  \\
\hline
Chunk-Conformer (repro) & Chunk-wise &$48$  &28.20  \\
SChunk-Conformer~\cite{Wang2022ShiftedCE} (repro)  &Chunk-wise &$48$ &$28.35$ \\
\hline
U2~\cite{Zhang2020UnifiedSA}(repro) &Time-restricted  &$36$ &$42.12$ \\
\hline
\textbf{SSCFormer}  &Chunk-wise &\textbf{48} &\textbf{28.28} \\
 \hline
\end{tabular}
\end{table}

\begin{table}[t]
\centering
\setlength{\tabcolsep}{2.4 mm}
\caption{ Comparisons with other models on AISHELL-1.
The chunk size and the right context, if any, determine the latency, as outlined in ~\cite{Zhang2020UnifiedSA}. The symbol $\triangle$ represents an additional latency incurred during rescoring, while $+2$ denotes the time required to simulate future features, as reported in ~\cite{An2022CUSIDECS}. }
\label{tab:other}
\begin{tabular}{|c|c|c|c|c|c|}
\hline
Model &Type &Latency &CER\\
Architecture & &(ms) &(\%)\\
\hline
     Sync-Transformer\cite{Tian2019SynchronousTF} &Chunk-wise &$400$ &$8.91$\\
     HS-DACS \cite{Li2021HeadSynchronousDF} &Chunk-wise &$1280$ &$6.80$ \\
     SCAMA\cite{Zhang2020StreamingCM} &Memory &$600$ &$7.39$ \\
     MMA-narrow\cite{Inaguma2020EnhancingMM}  &Memory &$960$ &$7.50$ \\
     MMA-wide\cite{Inaguma2020EnhancingMM} &Memory &$1920$ &$6.60$ \\
     SChunk-Transformer~\cite{Wang2022ShiftedCE} &Chunk-wise &$640$+$\triangle$ &$6.43$ \\ 
     SChunk-Conformer~\cite{Wang2022ShiftedCE} &Chunk-wise &$640$+$\triangle$ &$5.77$ \\  SChunk-Conformer~\cite{Wang2022ShiftedCE}(repro) &Chunk-wise &$640$+$\triangle$ &$5.75$ \\  
     U2~\cite{Zhang2020UnifiedSA} &Time-restricted &$640$+$\triangle$ &$5.42$ \\
     U2~\cite{Zhang2020UnifiedSA}(repro) &Time-restricted &$640$+$\triangle$ &$5.45$ \\
     CUSIDE~\cite{An2022CUSIDECS}  &Time-restricted &$400$+$2$ &$5.47$ \\
    \textbf{SSCFormer (\#chunk size = 10)} &Chunk-wise &$400$+$\triangle$ &\textbf{5.76} \\
    \textbf{SSCFormer} &Chunk-wise &$640$+$\triangle$ &\textbf{5.33} \\
\hline
     U2++~\cite{Wu2021U2UT} &Time-restricted &$640$+$\triangle$ &$5.05$ \\
     U2++~\cite{Wu2021U2UT}(repro) &Time-restricted &$640$+$\triangle$ &$5.12$ \\
    \textbf{SSCFormer (w/ bidecoder~\cite{Wu2021U2UT})} &Chunk-wise &$640$+$\triangle$ &\textbf{5.10} \\
\hline
     WNARS(w/ LM)~\cite{Wang2021WNARSWB} &Time-restricted &$640$+$\triangle$  &$5.22$ \\
     U2++(w/ LM)~\cite{Wu2021U2UT} &Time-restricted &$640$+$\triangle$ &$4.75$ \\
     CUSIDE(w/ LM)~\cite{An2022CUSIDECS}  &Time-restricted &$400$+$2$ &$4.79$ \\
    \textbf{SSCFormer (w/ LM)} &Chunk-wise &$640$+$\triangle$ &\textbf{4.78} \\
\hline
\end{tabular}
\end{table}

\textbf{Training and inference efficiency comparison.} In Table~\ref{tab:baseline} and Figure~\ref{fig:time_cost}, we provide a comparison of the training and inference efficiency of Chunk-Conformer, SChunk-Conformer~\cite{Wang2022ShiftedCE}, U2~\cite{Zhang2020UnifiedSA}, and SSCFormer.
We can see that SSCFormer can achieve similar efficiency as the other chunk-wise models both in training and inference stages.
When comparing with U2, which is a strong time-restricted model, we observe that our proposed model, SSCFormer, can train with larger batch sizes, leading to significantly reduced training time. Furthermore, the Real Time Factor (RTF) of our model is insensitive to the audio length, whereas U2 has a quadratic RTF.
Indeed, the ability of SSCFormer to maintain a consistent RTF, regardless of audio length, can be beneficial in controlling latency in real-world systems.

\textbf{Performance comparison.} As previously discussed, the SSCFormer architecture offers benefits in terms of both training and inference efficiency.
Here, we present a comparison of the SSCFormer's performance in terms of CER with other state-of-the-art (SOTA) models, including several strong time-restricted baseline models.
Table~\ref{tab:other} demonstrates that the SSCFormer outperforms other chunk-wise or memory-based models by a significant margin (7.3\%+ relative CER reduction), achieving a CER of 5.33. Furthermore, it even outperforms recently released time-restricted models, such as U2\cite{Zhang2020UnifiedSA} and CUSIDE\cite{An2022CUSIDECS} by 2\%+ relative CER reduction.
E2E ASR models can be further enhanced using other advanced techniques such as a bi-directional Attention decoder (bidecoder) or a Language Model (LM) to rescore the final outputs. Similarly, the performance of the SSCFormer can also be improved by incorporating these techniques, enabling it to achieve competitive CERs compared to SOTA models.

\section{Conclusion}
\label{sec:conclusion}
Our findings challenge the traditional streaming schemes used in online Conformer models. The SSCFormer improves global context modeling using a sequentially sampled chunk scheme and integrates chunked convolution and causal convolution to capture chunk-wise future information in convolution layers. It outperforms current chunk-wise models and achieves comparable or superior results to existing quadratic models. Moreover, our approach is not limited to specific network architectures, making it a valuable addition for streaming deployment in other ASR models, such as Squeezeformer \cite{Kim2022SqueezeformerAE}.

\bibliographystyle{IEEEtran}
\bibliography{checkpoint.bib}
\end{document}